\documentstyle{mn}

\input epsf

\begin{document}
\journal{Preprint astro-ph/0011187} 
\title[The kinetic Sunyaev--Zel'dovich effect] 
{Hydrodynamical simulations of the Sunyaev--Zel'dovich effect: the kinetic 
effect} 
\author[A.~C.~da Silva et al.]{Antonio C.~da Silva,$^1$
Domingos Barbosa,$^{2,3}$ Andrew R.~Liddle$^1$\cr and
Peter A.~Thomas$^1$\\ 
$^1$Astronomy Centre, University of Sussex, Brighton BN1 9QJ\\ 
$^2$Astrophysics Group, Lawrence Berkeley Laboratory, 1 Cyclotron Road, 
Berkeley CA 94710, USA\\
$^3$CENTRA, Instituto Superior Tecnico, Av. Rovisco Pais, 1049-001 Lisboa
Codex, Portugal}

\maketitle

\begin{abstract}
We use hydrodynamical $N$-body simulations to study the kinetic
Sunyaev--Zel'dovich effect. We construct sets of maps, one square
degree in size, in three different cosmological models. We confirm
earlier calculations that on the scales studied the kinetic effect is
much smaller than the thermal (except close to the thermal null
point), with an rms dispersion smaller by about a factor five
in the Rayleigh--Jeans region. We study the redshift dependence of the
rms distortion and the pixel distribution at the present epoch. We
compute the angular power spectra of the maps, including their redshift 
dependence, and compare them with
the thermal Sunyaev--Zel'dovich effect and with the expected cosmic
microwave background anisotropy spectrum as well as with
determinations by other authors. We correlate the kinetic
effect with the thermal effect both pixel-by-pixel and for identified
thermal sources in the maps to assess the extent to which the kinetic effect is 
enhanced in locations of strong thermal signal.
\end{abstract}



\section{Introduction}

The Sunyaev--Zel'dovich (SZ) effect (Sunyaev \& Zel'dovich 1972, 1980;
for reviews see Rephaeli 1995 and Birkinshaw 1999) is the change in
energy experienced by cosmic microwave background photons when they
scatter from intervening gas, especially that in galaxy clusters. The
dominant version from clusters is the thermal SZ effect, the gain in
energy acquired from the thermal motion of the gas which is commonly
at a temperature of tens of millions of degrees in clusters. The
kinetic SZ effect is the Doppler shift arising from the bulk motion of
the gas.

The thermal effect has been quite well studied theoretically, and
recently has become a burgeoning area of observational activity with
the construction of two-dimensional maps of clusters becoming
commonplace (Jones et al.~1993; Myers et al.~1997; Carlstrom et
al.~2000).  By contrast, less attention has been given to the kinetic
effect.  It presents a considerable observational challenge, because
it is predicted to be much smaller than the thermal effect on the
angular scales explored so far, and also because unlike the thermal
effect it possesses no characteristic spectral signature allowing it
to be distinguished from primary cosmic microwave background
anisotropies.  It is also more difficult to make semi-analytic
calculations. The thermal effect can be obtained fairly directly via
the Press--Schechter (1974) approach; various aspects including number counts, 
the global CMB distortion, and its 
impact as secondary anisotropy for CMB measurements at small scales 
have been extensively studied (Cole \& Kaiser~1988; Colafrancesco et
al.~1994, 1997; Bartlett \& Silk~1994; Barbosa et al.~1996; Eke, Cole \& 
Frenk 1996; Aghanim et al.~1997; Komatsu \& Kitayama 1999; 
Atrio-Barandela \& M\"ucket 1999; Molnar \& Birkinshaw 2000; 
Hern\'andez-Monteagudo, Atrio-Barandela \& M\"ucket 2000). 
Predictions for the kinetic effect require the simultaneous estimation of both
mass and peculiar velocity, and the signal is much less dominated by the gas
which happens to be in massive halos.  A detailed analytical model has
recently been constructed by Valageas, Balbi \& Silk (2000) [see also
Benson et al.~2000], which computes the effect including
inhomogeneities in reionization; this supersedes earlier modelling by
Aghanim et al.~(1997) where the peculiar velocities were drawn from a
gaussian distribution of fixed width.

We recently used large-scale hydrodynamical $N$-body simulations to
make simulated maps of the thermal effect and analyzed their
properties (da Silva et al.~2000; for similar recent work see
Refregier et al.~2000a; Seljak, Burwell \& Pen 2000; Springel, White \&
Hernquist~2000).  In this paper, we make a detailed analysis of maps
of the kinetic effect, made using the same technique.  Recently
Springel et al.~(2000) used similar simulations to study the angular
power spectrum of the kinetic effect in the $\Lambda$CDM cosmology,
and investigated the influence of non-gravitational heating upon it.
In this paper we study three different cosmologies, and investigate a
wide range of properties of the resulting maps including the relation
between thermal and kinetic distortion.

\section{The kinetic SZ effect}

\subsection{Linear and non-linear effects}

Traditionally, the phrase ``kinetic SZ effect'' is associated with the
scattering of photons from galaxy clusters or other large
gravitationally-bound objects.  However, in simulations there is no
clear distinction between this and the scattering from any other
moving gas, and when we construct maps they will be based on the bulk
motions of all the material in the simulation box.  However, one
should recognize that part of this signal is computable in linear
perturbation theory, and is for example computed in a run of the {\sc
cmbfast} program (Seljak \& Zaldarriaga 1996) for a cosmology with
reionization, where it is called the low-redshift Doppler effect.
These different regimes, and their relation to the Ostriker--Vishniac
effect (Ostriker \& Vishniac 1986), have been discussed in detailed by
Hu (2000) and by Gnedin \& Jaffe (2000), and observational aspects are
described by Cooray \& Hu (2000).

Concerning the contribution from non-linear evolution, one should
recognize that while there are non-linear contributions to the
velocities, of particular importance is the non-linear evolution of
the density field.  This concentrates the matter into a virialized
region which gives a high optical depth probe of the bulk velocity at
that point.  The non-linear effect is more important at low redshifts.

\subsection{The reionization epoch}

Photons only scatter at epochs where the gas is ionized.  Reionization
is thought to be induced by radiation from the first generation of
massive stars in dwarf galaxies (see Haiman \& Knox 1999 for a
review), which unfortunately are well below our numerical resolution.
Bruscoli et al.~(1999) and Gnedin \& Jaffe (2000) have modelled
reionization using very small-scale high-resolution simulations,
probing a range of angular scales complementary to our own, and
compute the corresponding microwave anisotropies.  We will not
consider inhomogeneous reionization in this paper (see Aghanim et
al.~1996; Grusinov \& Hu 1998; Knox, Scoccimarro \& Dodelson 1998;
Benson et al.~2000; Valageas et al.~2000), and we will treat the epoch of 
reionization as an
independent parameter. According to Gnedin \& Jaffe (2000),
anisotropies from inhomogeneous reionization will be subdominant on
the scales we consider, especially if massive stars are indeed
responsible for reionization.

For the thermal SZ effect the epoch of reionization is irrelevant, as
significant signals only come from low redshifts and from regions hot
enough to be fully ionized. This is no longer true for the kinetic
effect, where the increased density of material at high redshift (per
observed solid angle) maintains a significant signal even though the
velocities are smaller. Indeed, we will see that, unlike the thermal
SZ signal, the kinetic SZ signal is not convergent and results are
mildly dependent on the assumed redshift of reionization (though this
is certainly not the largest uncertainty in our calculations). For
ease of comparison between cosmologies, we will assume that
reionization took place at redshift 10, which satisfies all existing 
observations (see e.g.~Griffiths et
al.~1999; Haiman \& Knox 1999), independent of the cosmology even though 
detailed
calculations would favour later reionization in the high-density case.

\subsection{The equations}

The thermal SZ effect in a given direction is computed as a line
integral, which gives the Compton $y$-parameter as (Sunyaev \&
Zel'dovich 1972, 1980)
\begin{equation}
\label{eq1}
y = \int {{k_{{\rm B}} \sigma_{{\rm T}}} \over {m_{{\rm e}} c^2}}\, \,
	T_{{\rm e}} \, n_{{\rm e}}\, dl \,.
\end{equation}
In this expression $T_{{\rm e}}$ and $n_{{\rm e}}$ are the temperature
and number density of the electrons, $k_{{\rm B}}$ the Boltzmann
constant, $\sigma_{{\rm T}}$ the Thomson cross-section, $c$ the speed
of light and $m_{{\rm e}}$ the electron rest mass. The frequency
dependence of the effect is well documented (e.g.~Birkinshaw 1999);
ignoring relativistic corrections, it gives a temperature fluctuation
\begin{equation}
\frac{\Delta T_{{\rm th}}}{T} = y \left[ \frac{x}{\tanh(x/2)} - 4 \right] \,.
\end{equation}
where $x = h\nu/k_{{\rm B}}T$ is the dimensionless frequency and $T$
the radiation temperature.  In the long-wavelength limit $x \ll 1$
(the Rayleigh--Jeans portion of the spectrum) we have $\Delta T_{{\rm
th}}/T \simeq -2y$, and the thermal effect has a null point at $x
\simeq 3.83$, corresponding to $\nu \simeq 217 \, {\rm GHz}$.

Whenever the gas has a peculiar velocity in the CMB radiation frame,
there is a kinematic Doppler effect in addition to heating from the
random thermal motions of the electrons. The change in the spectral
intensity due to the kinetic effect is (Sunyaev \& Zel'dovich 1980)
\begin{equation}
\Delta I_{\nu }=-I_0\,h(x)\,{{v_{\rm r}} \over c}\, \tau
\end{equation}
where $I_0=2 k_{{\rm B}}^3 T^3/h^2 c^2$, $v_{{\rm r}}$ is the
line-of-sight velocity of the gas (positive/negative when
receding/approaching), $h(x)$ is given by
\begin{equation}
h(x)={{x^4e^x} \over {\left( {e^x-1} \right)^2}} \,,
\end{equation} 
and $\tau $ is the optical depth
\begin{equation}
\tau =\int {\sigma _{{\rm T}}\, n_{{\rm e}}\, dl} \,.
\end{equation} 
The temperature fluctuation due to the kinetic effect is 
\begin{equation}
\label{eq2}
{{\Delta T_{{\rm k}}} \over {T}}=-{{v_{\rm r}} \over c}\,\tau=-\int
{{{v_{{\rm r}}} \over c}\,\sigma _{{\rm T}}\, n_{{\rm e}}\, dl} \,,
\end{equation}
which, in contrast to the thermal effect, is a frequency-independent
quantity.  As such, the kinetic effect cannot be distinguished from
primary microwave anisotropies on the basis of its spectral
dependence.

\section{The maps}

For full details as to the construction of maps, see da Silva et
al.~(2000). The starting ingredient is a set of three hydrodynamical
simulations, one for each of three cosmologies.
\begin{itemize} 
\item $\Lambda$CDM: a low-density model with a flat spatial geometry and 
$\Omega_0 = 0.35$ and $\Omega_\Lambda = 0.65$.
\item $\tau$CDM: a critical-density model with $\Omega_0 = 1$ and 
\mbox{$\Omega_\Lambda = 0$}.
\item OCDM: a low-density open model with $\Omega_0 = 0.35$ and
$\Omega_\Lambda = 0$.
\end{itemize}
The simulations were carried out using the public domain Hydra code
(Adaptive P$^3$M-SPH: Couchman, Thomas \& Pearce 1995).  In each case
the box-size was $150\,h^{-1}$ Mpc, and equal numbers of dark matter
and gas particles were used.

To construct a map we stack simulation output boxes, with random
translations and reorientations, to reach high redshift.  Outputs from
the simulations were made at different epochs so that they can be
stacked smoothly into an evolving sequence.  Typically around 35 boxes
need to be stacked to reach to redshift 10; however at the redshifts from which 
the bulk of the signal arises only a small fraction of the volume of the box
contributes to the maps and so we are not limited by cosmic variance
in having only a single hydrodynamical simulation.  For each cosmology
we make 30 different maps, corresponding to different random
orientations of the stacked simulations.

To compute the thermal and kinetic effects we follow the approach
described in da Silva et al.~(2000). In the simulations each gas
particle occupies a volume with radius proportional to its SPH
smoothing radius, $h_{\rm i}$, inside which the mass profile is taken
as $m_{{\rm gas}}W({\bf r}-{\bf r}_{\rm i}, h_{\rm i})$, where ${\bf
r}_{\rm i}$ is the position of the particle's centre, $m_{{\rm gas}}$
is the mass of the gas particle and $W$ is the normalized
spherically--symmetric smoothing kernel.

The quantities we are interested in mapping are $y$ and $\Delta
T_{{\rm k}}/T$.  To do this we first convert the line-of-sight
integrations in equations ({\ref{eq1}) and (\ref{eq2}) into volume
integrals over space.  The resulting expressions are then discretized
into cubic cells of volume V (voxels), which are stacked along the
line of sight to produce pixels of area $A$. The total temperature
distortion induced by the kinetic effect in each pixel of the map can
then be evaluated as
\begin{equation}
\label{eq3}
{{\Delta T_{{\rm k}}} \over {T}} = -\frac{\sigma_{{\rm T}}}{c} \,
\frac{V}{A} \, {0.88 \, m_{{\rm gas}} \over {m_{{\rm p}}}}
\sum\limits_{\rm i}\sum\limits_{\alpha }v_{{\rm r},i} \, W(\left|
{{\bf r_{\alpha}}-{\bf r}_{\rm i}} \right| ,h_{\rm i})\,,
\end{equation}
and likewise for the thermal SZ $y$-parameter 
\begin{equation}
\label{eq4}
y={k_{{\rm B}} \sigma_{{\rm T}} \over {m_{{\rm e}} c^2}} \, 
  \frac{V}{A} \, {0.88 \, m_{{\rm gas}} \over {m_{{\rm p}}}}
  \sum\limits_{\rm i}\sum\limits_{\alpha } T_{\rm i} \, W(\left|
  {{\bf r_{\alpha }}-{\bf r}_{\rm i}} \right| ,h_{\rm i}) \,,
\end{equation}
where the $i$ index runs over all particles which contribute to the
pixel columns, and the $\alpha$ index is over the line-of-sight
cubes. For each particle the sum of the smoothing kernel over voxels
is normalized to $1/V$. The quantities $y$ and $\Delta T_{{\rm k}}/T$
are the average values in each pixel. In these expressions, $v_{{\rm
r},i}$ and $T_{\rm i}$ are the velocity and the temperature of the gas
particles, $m_{{\rm p}}$ is the proton mass, and the factor 0.88 gives
the number of electrons per baryon, assuming a 24 per cent helium
fraction and complete ionization in the regions of interest. We
construct maps of the individual simulation boxes in our stacking,
which are then added to give the final maps.

\begin{figure*}
\centering \leavevmode\epsfysize=7cm \epsfbox{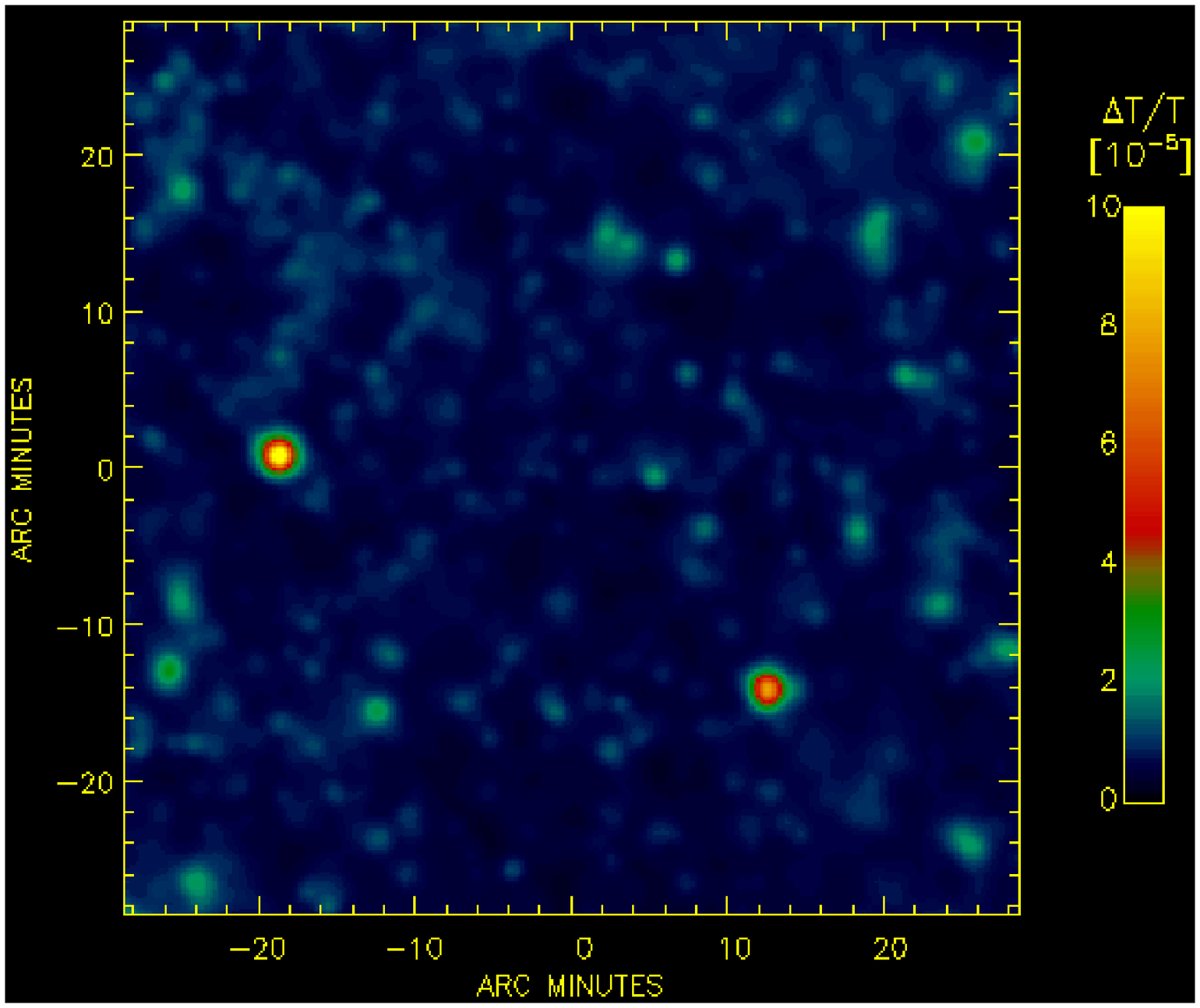} 
\hspace*{0.15cm} \leavevmode\epsfysize=7cm 
\epsfbox{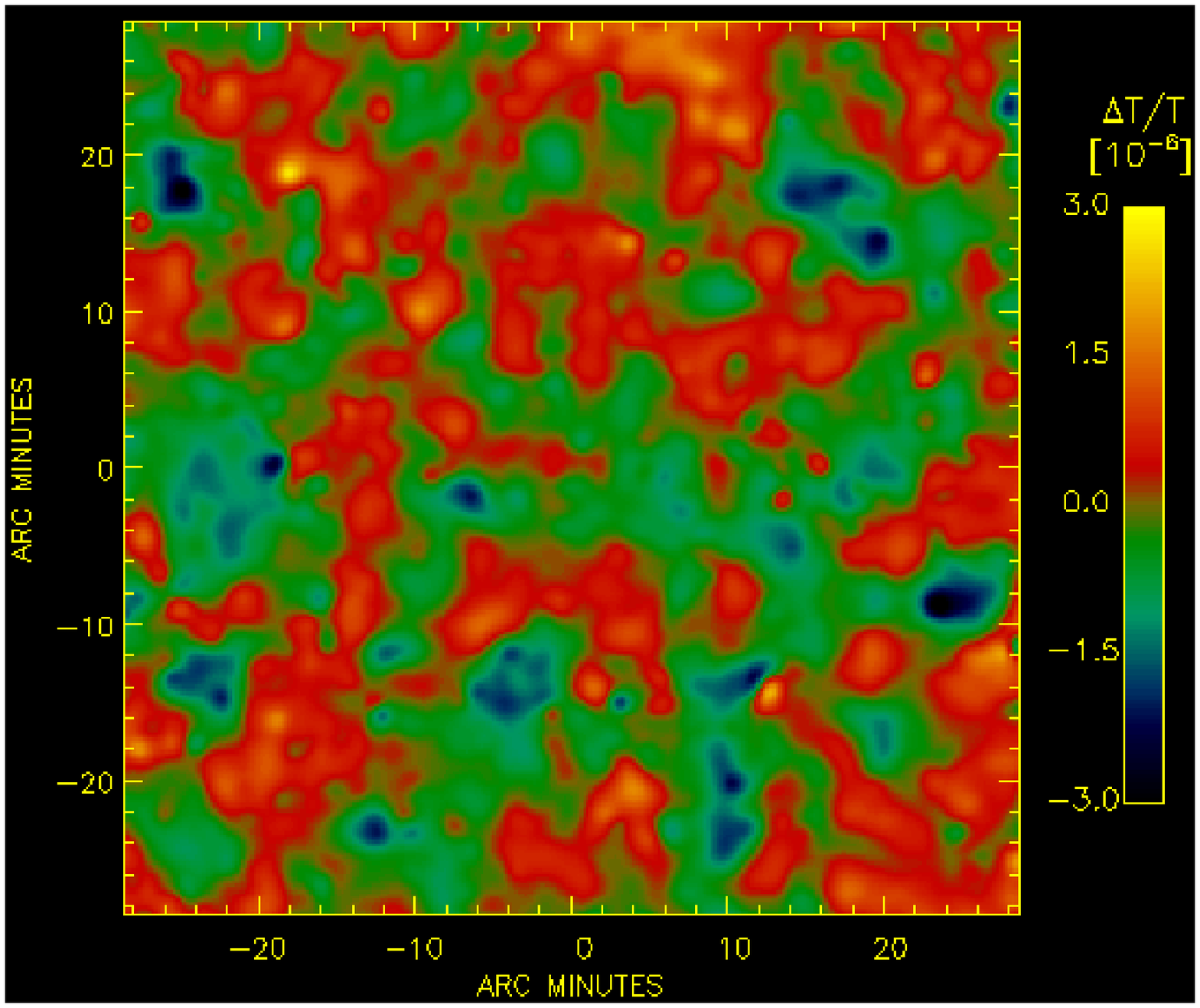} 
\vspace*{0.15cm} 
\centering \leavevmode\epsfysize=7cm \epsfbox{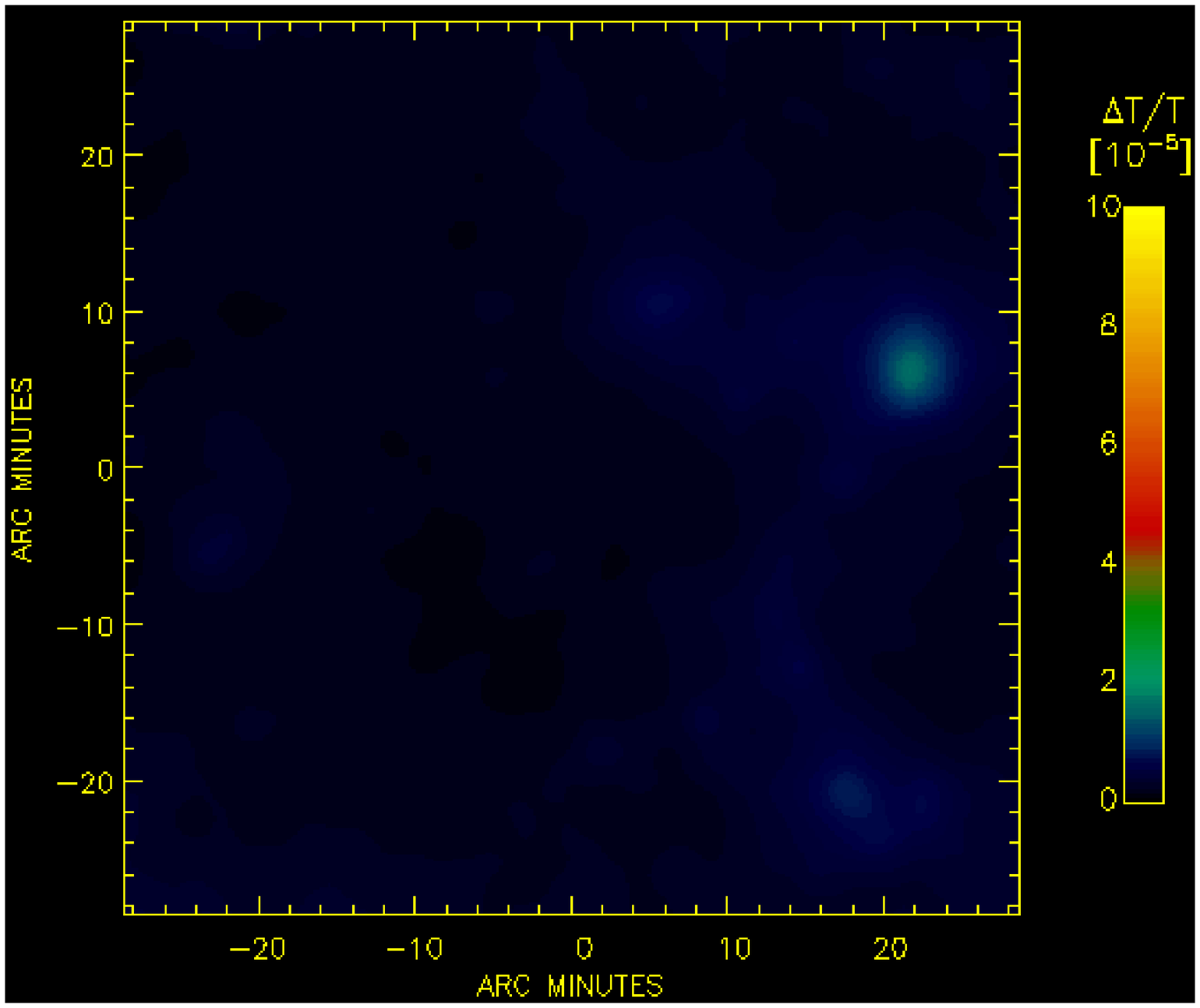} 
\hspace*{0.15cm} \leavevmode\epsfysize=7cm 
\epsfbox{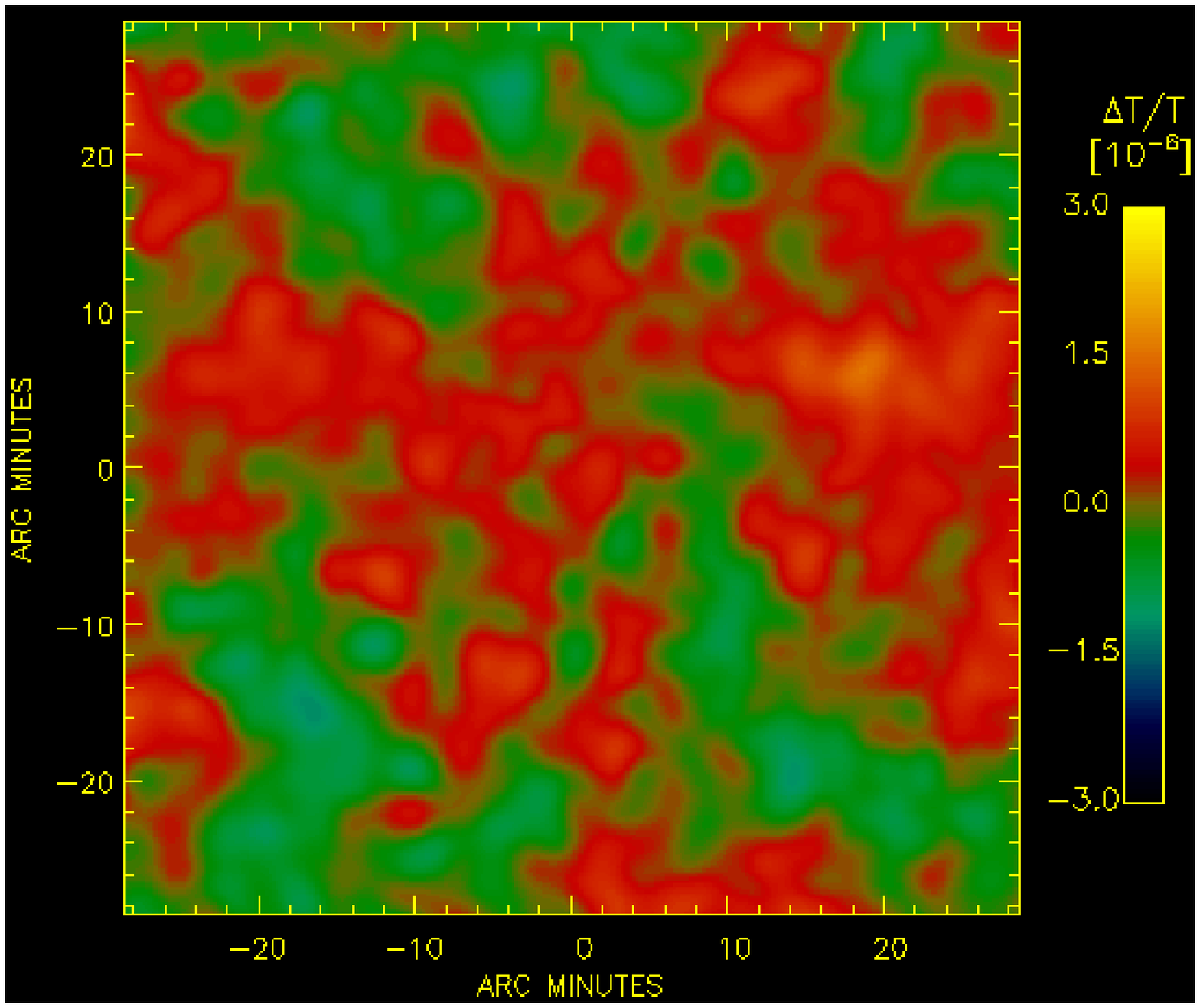}\\ 
\vspace*{0.15cm}
\centering \leavevmode\epsfysize=7cm \epsfbox{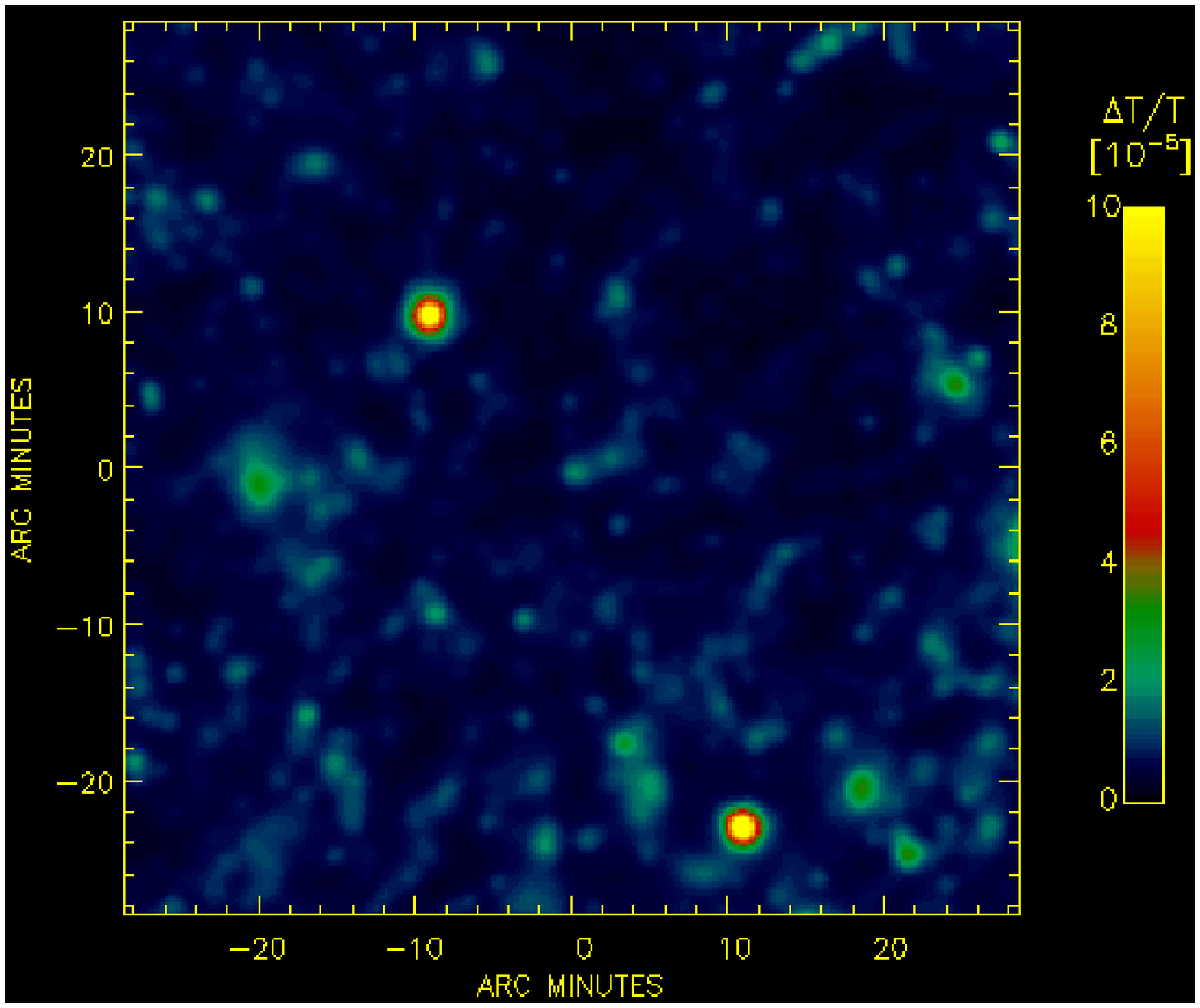} 
\hspace*{0.15cm} \leavevmode\epsfysize=7cm 
\epsfbox{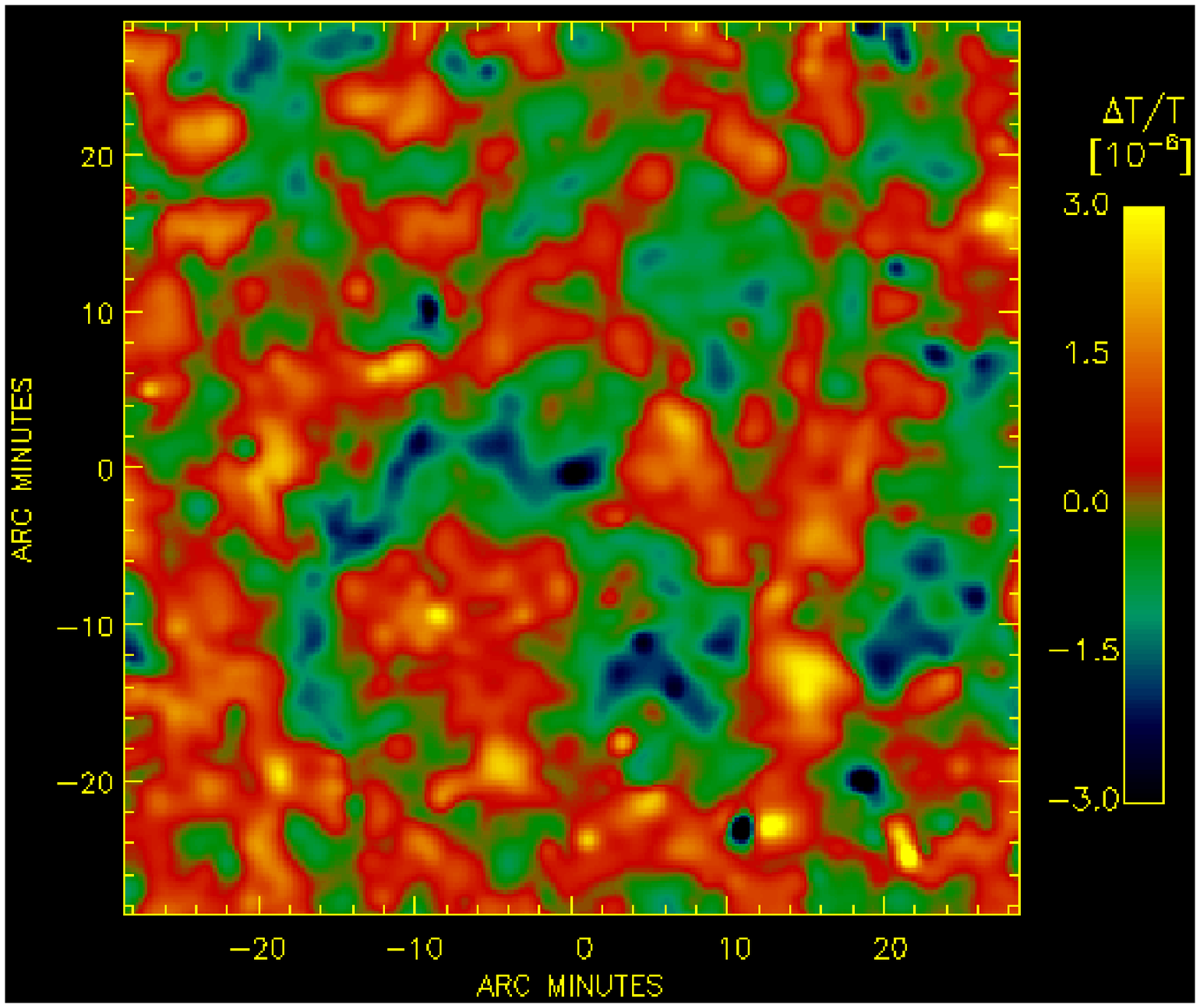}\\ 
\caption[Figure1]{\label{f:maps} Pairs of simulated SZ maps of size
one square degree. The left panels show $\Delta T_{{\rm th}}/T$ in the
Rayleigh--Jeans part of the spectrum, while the right panels show the
corresponding kinetic maps of $\Delta T_{{\rm k}}/T$ from the same map
realization. In each case, the original map was convolved with a
gaussian beam profile of FWHM = 1\arcmin.
From top to bottom, the three sets are $\Lambda$CDM, $\tau$CDM and
OCDM.}
\end{figure*}

\section{Results}

Typical example maps in each cosmology are shown in Figure~\ref{f:maps}, with
the same colour scale in each.\footnote{A more extensive selection of colour
maps, along with animations showing the contribution from each redshift, can be
found at {\tt star-www.cpes.susx.ac.uk/$\sim$andrewl/sz/sz.html}} The original
maps have been smoothed using a gaussian with a full-width half-maximum (FWHM)
of 1\arcmin.  Unlike the thermal maps, the kinetic maps show
positive and negative temperature fluctuations.  The low-density maps (top and
bottom panels) present stronger fluctuations than the critical density ones
(centre panels).  The brightest objects in the thermal maps do not necessarily
correspond to the largest kinetic distortions (and vice-versa).  However, bright
thermal sources often reveal neighboring kinetic peaks of opposite signs 
(Diaferio,
Sunyaev \& Nusser 2000), 
resulting from
mergings of substructures or rotational flows within the sources.  This can be 
seen, for example, corresponding to the
two brightest objects of the thermal $\Lambda $CDM map and in the bright source
on the lower right corner of the OCDM map.  These objects are at redshifts
z=1.7, 1.8, and 1.9 respectively.  Fainter super-clusters at lower redshifts
also produce similar signatures (see animations of individual maps on our WWW
page), but their contribution to the final map tends to be erased by the box
stacking process.  Although the kinetic effect has no spectral signature, the
search for these characteristic positive/negative features in the SZ flux of
bright sources and nearby super-clusters may prove to be an additional aid in
isolating the kinetic effect in future SZ measurements and in identifying
super-clusters at cosmological distances (see Diaferio et al.~2000).

\begin{figure}
\centering \leavevmode\epsfysize=11cm \epsfbox{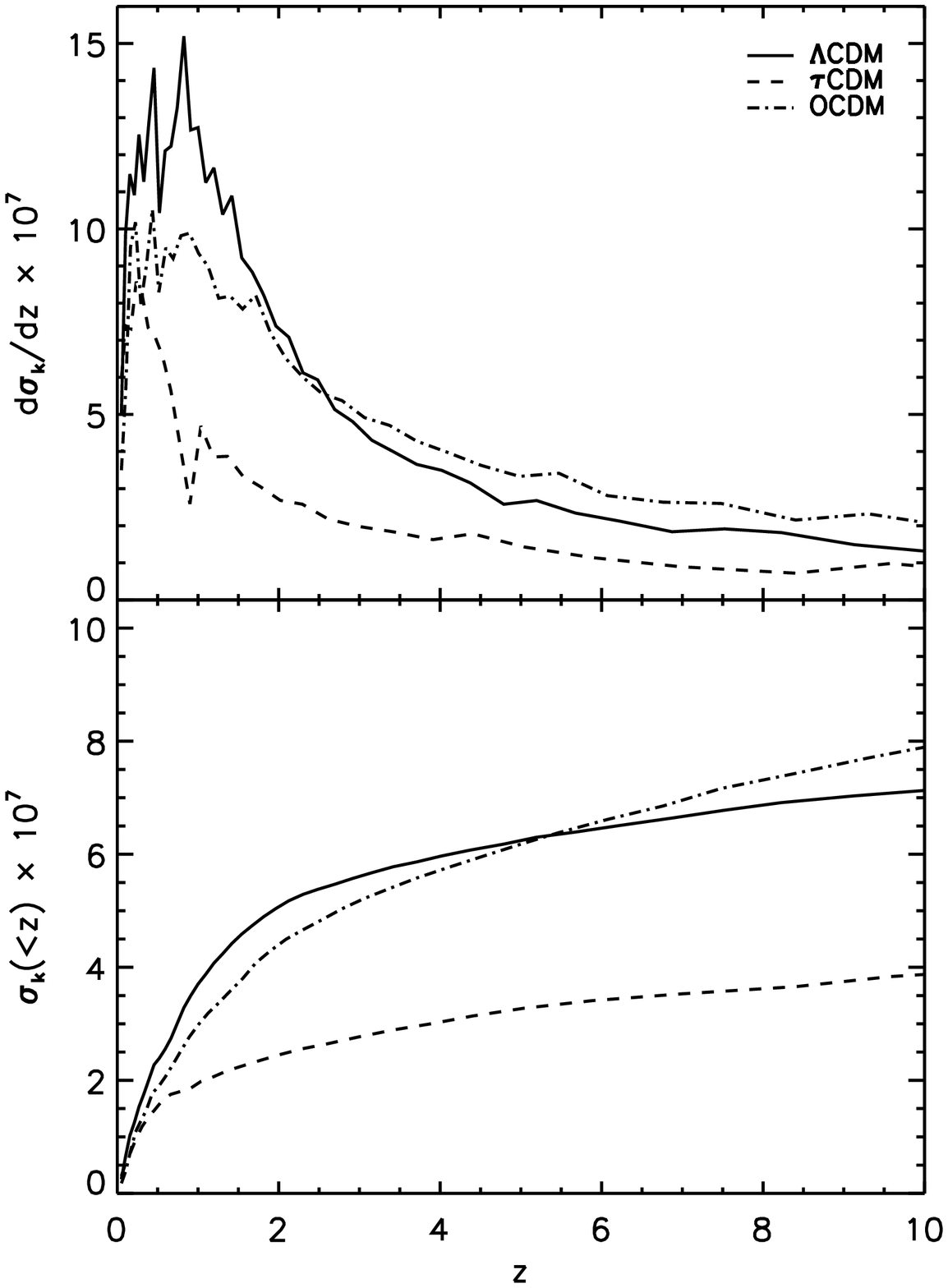}\\
\caption[dkdz_kz_all]{\label{f:dkdz_kz_all} The rms kinetic distortion
in the 1\arcmin~resolution maps made from individual simulation boxes at a
redshift $z$ (upper panel), and the rms of the accumulation of these
maps up to redshift $z$ (lower panel).  Each curve is an average over
the 30 map realizations. Note that the lower panel is not simply the
integral of the upper, because when kinetic maps are added together they
`interfere'.}
\end{figure}

\begin{figure}
\centering \leavevmode\epsfysize=11cm \epsfbox{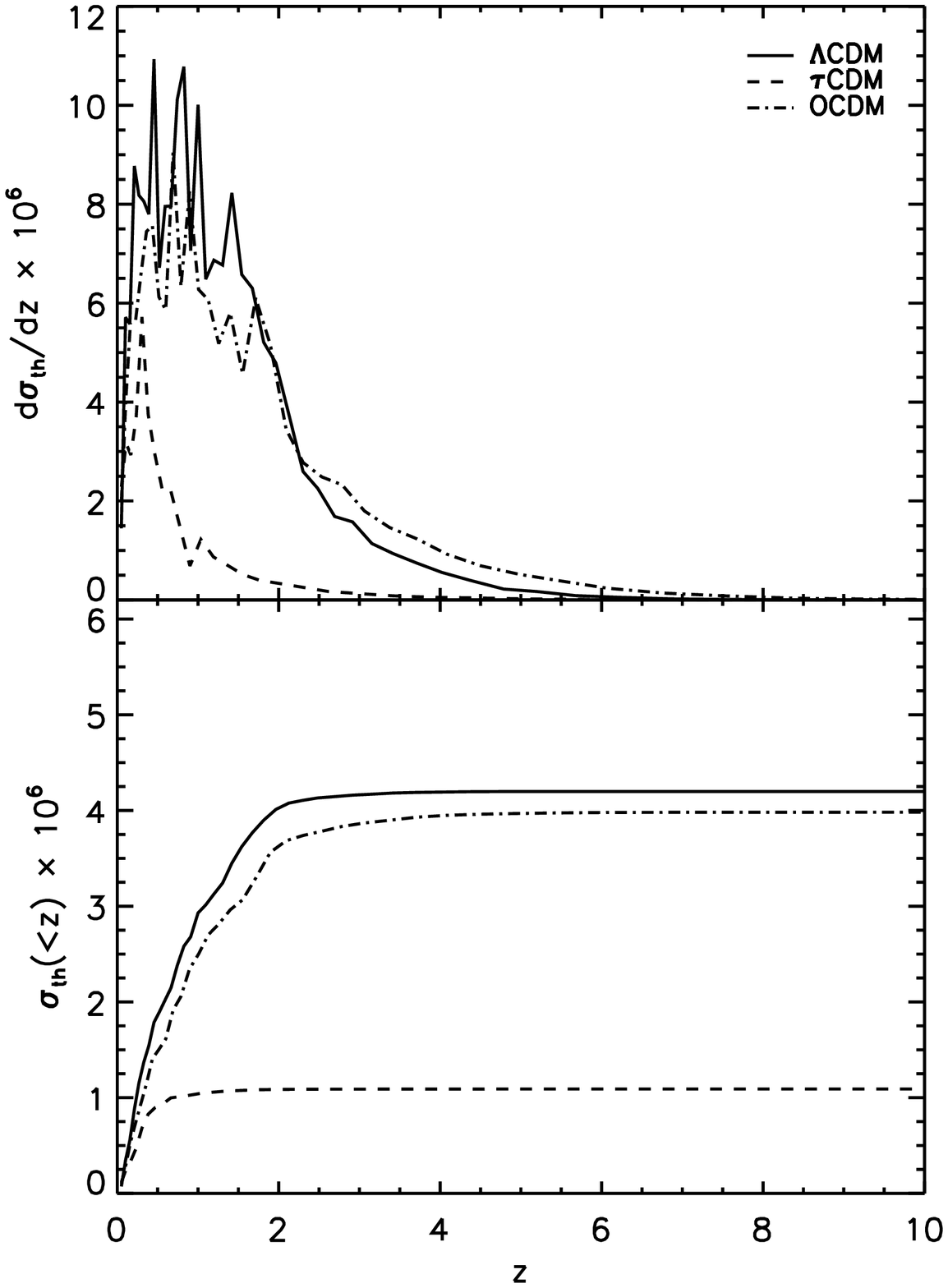}\\ 
\caption[dydz_yz_all]{\label{f:dydz_yz_all} As
Figure~\ref{f:dkdz_kz_all} but for the thermal effect. Strong
convergence to high redshift is seen. Note that this plot, showing the
dispersion, differs from a similar plot in da Silva et al.~(2000)
which showed the convergence of the mean $y$-distortion with
redshift.}
\end{figure}

\subsection{The redshift dependence}

In order to test how well the calculation of the effects converges with
redshift, Figure~\ref{f:dkdz_kz_all} shows the rms distortion, $\sigma
_{\rm k}=\sqrt{\left< (\Delta T_{\rm k}/T)^2 \right> }$, in the 1
arcminute resolution maps constructed out to redshift $z$, as a
function of that redshift. Unlike the dispersion of the thermal
effect, shown in Figure~\ref{f:dydz_yz_all}, the integrated kinetic
signal is not strongly convergent to high redshift, with the smaller
velocities being compensated by the increased amount of material
contributing. In the real Universe the signal is cut off by the epoch
of reionization, and as can be seen our results are modestly dependent
on that choice (though other uncertainties such as the normalization
of the matter power spectrum will be more significant). As advertised
in Section~2, we choose to construct maps with data out to redshift 10
in each cosmology. 

\subsection{The dispersion}

In Table~1 we list the rms fluctuations in both the thermal and kinetic effects, 
for a range of different smoothings to represent possible instrumental 
resolutions. They are given as temperature fluctuations in the long-wavelength 
part of the spectrum. Because the thermal effect is purely additive, its 
fluctuations, which is all that most instruments are capable of measuring, do 
not dominate over the kinetic fluctuations by as large a factor as one would 
surmise from looking at the total magnitude of the effects.

\begin{table}
\centering
\caption{The rms fluctuations of the thermal ($\sigma_{{\rm th}}$) and kinetic 
($\sigma_{{\rm k}}$) SZ
effects as a function of beam resolution in the Rayleigh--Jeans limit. Values 
are in multiples of $10^{-7}$.}
\label{tab:rms}
\begin{tabular}{|r|cc|cc|cc|cc|cc|}
\hline
 & \multicolumn{2}{|c|}{1'} &
 \multicolumn{2}{c}{2'} & \multicolumn{2}{|c|}{5'}&\multicolumn{2}{c}{10'} \\
 & $\sigma_{{\rm th}} $ & $\sigma_{{\rm k}} $ & $\sigma_{{\rm th}} $ & 
$\sigma_{{\rm k}} $ &
 $\sigma_{{\rm th}} $ & $\sigma_{{\rm k}} $ &$\sigma_{{\rm th}} $ & 
$\sigma_{{\rm k}} $  \\
\hline
$\Lambda $CDM & 42 & 7.2 & 33 & 6.1 & 20 & 4.1 & 11 & 2.4 \\
$\tau $CDM & 11 & 3.8 & 10 & 3.5 & 8.5 & 2.6 & 5.9 & 1.7  \\
OCDM & 40 & 8.0 & 31 & 6.7 & 18 & 4.4 & 10 & 2.5 \\
\hline
\end{tabular}
\end{table}

\subsection{The pixel distribution}

Figure~\ref{f:histogram} shows a histogram of the pixel distribution
in the kinetic maps for the $\Lambda$CDM cosmology. The different
curves represent different smoothings ranging from 0.5\arcmin\ to
10\arcmin.  The curves show the distribution of $\Delta T_{\rm k} /T$
values which would be seen if an instrument with the corresponding
resolution points randomly at the sky. At low resolutions (5
arcminutes or worse) the curves are close to gaussian distributed for
all intensities, whereas at higher resolution marked nongaussian tails
are present from non-linear structures.  These nonlinear tails
represent the best hope for observing the effect directly.  As
expected the pixel distribution mean values are very close to zero.

These pixel distributions are quite discouraging for attempts to
detect the kinetic effect from individual sources with the {\it Planck}
satellite, whose best resolution is about five arcminutes. We see that
the pixel distribution at such resolution is well approximated by a
gaussian even for the rarest events we see, with no non-gaussian tail
of bright rare events. The rms dispersion is very small compared to
the expected level of primary anisotropies on those scales, and so the
kinetic effect will not be detectable directly without some external
information as to where to look to see it. Later in this paper, we
shall consider the use of the thermal effect as an indicator of where
to look for the kinetic effect. This can allow a statistical measure
of the cluster velocity field (Aghanim et al.~1997; see also Haehnelt
\& Tegmark 1996; Hobson et al.~1998). Indeed, the MAP and {\it Planck} missions 
could allow
the overall detection of the cosmic bulk flow when coupled with X-ray 
observations (Kashlinsky \& Atrio-Barandela 2000). 
 
\subsection{The angular power spectrum}

We calculate the power spectrum using the flat-sky approximation in
the usual way; see e.g.~Peacock (1999).
In Figure~\ref{f:cls} we show
the thermal and kinetic SZ angular power spectra obtained from
simulations for the three cosmologies, with the $C_{\ell}$
band averaged into logarithmic bins in the Fourier space.

\begin{figure}
\centering \leavevmode\epsfysize=6cm \epsfbox{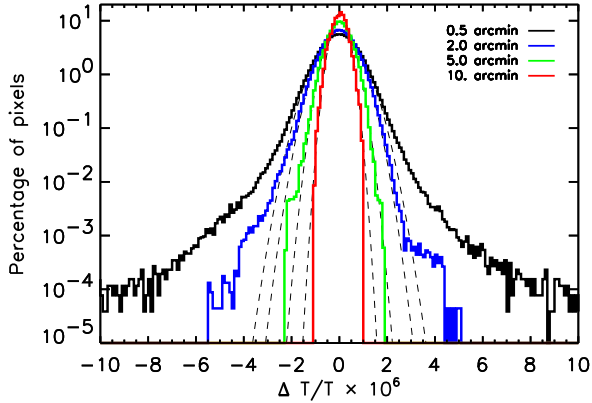}\\
\caption[histogram]{\label{f:histogram} A histogram of the $\Delta
T_{{\rm k}}/T$ values in the $\Lambda$CDM maps with different smoothings. The
dashed lines show gaussian fits to the distributions.}
\end{figure}

The power spectra shown are averages over the thirty maps we made for
each cosmology. To estimate the statistical error we used bootstrap
resampling of the thirty maps, which gave a range of estimates
indicated at 68 per cent confidence by the dotted lines (note that the
map-to-map variations are of course much bigger than the variation in
the bootstrap resamples of thirty maps each). For the two low-density
cosmologies we see that the statistical errors are small, while for
the critical-density thermal case even thirty maps leaves a
significant residual error, because the maps tend to be
dominated by the presense or absence of single bright
features. Unfortunately computer time limitations prevent us
increasing the number of maps used to make the estimate, and from
running more sets of hydrodynamical simulations to remove possible
correlations from using a single set.

\vspace{0.5cm}
\begin{figure}
\centering \leavevmode\epsfysize=6cm \epsfbox{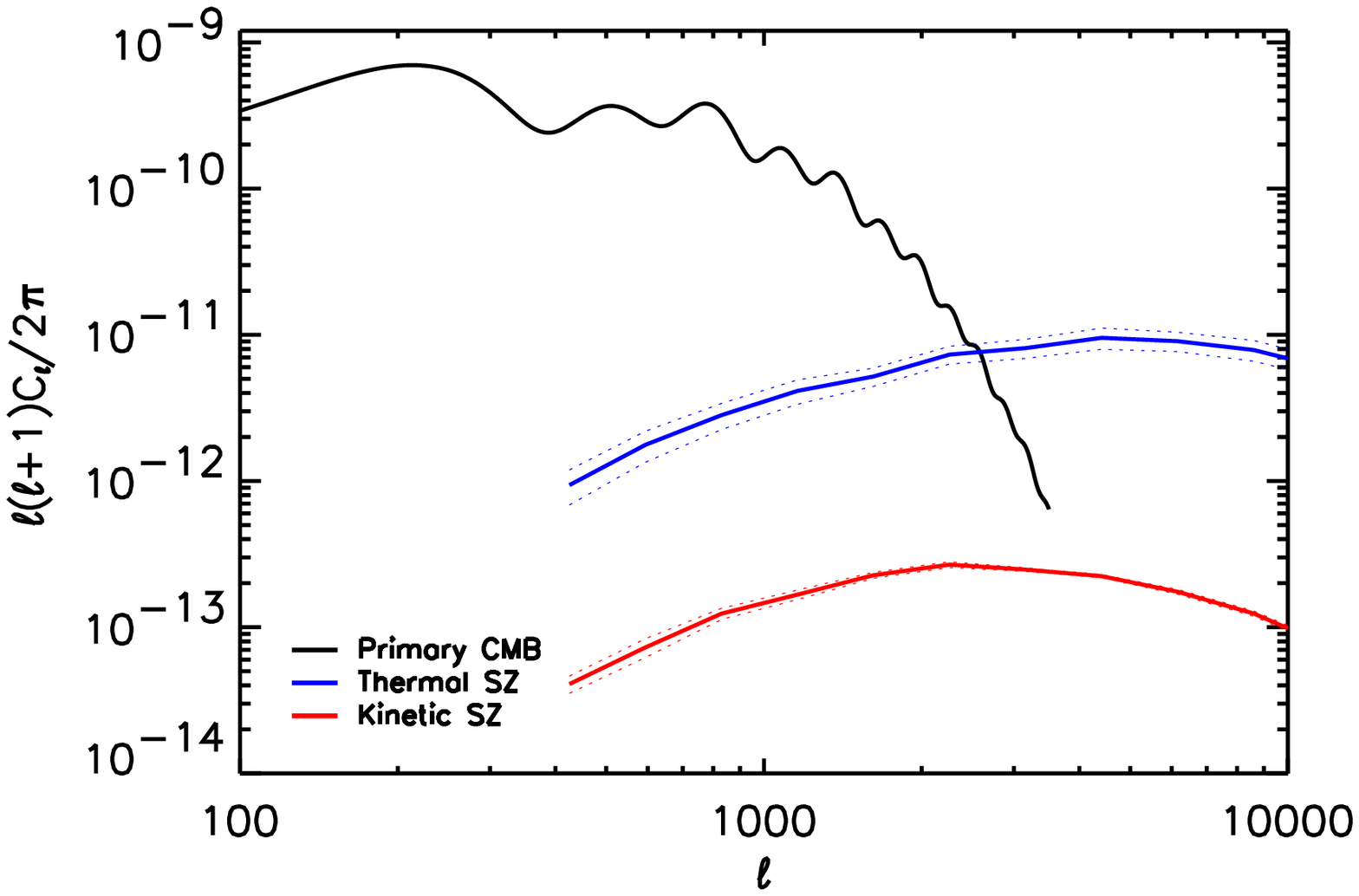}\\ 
\centering \leavevmode\epsfysize=6cm \epsfbox{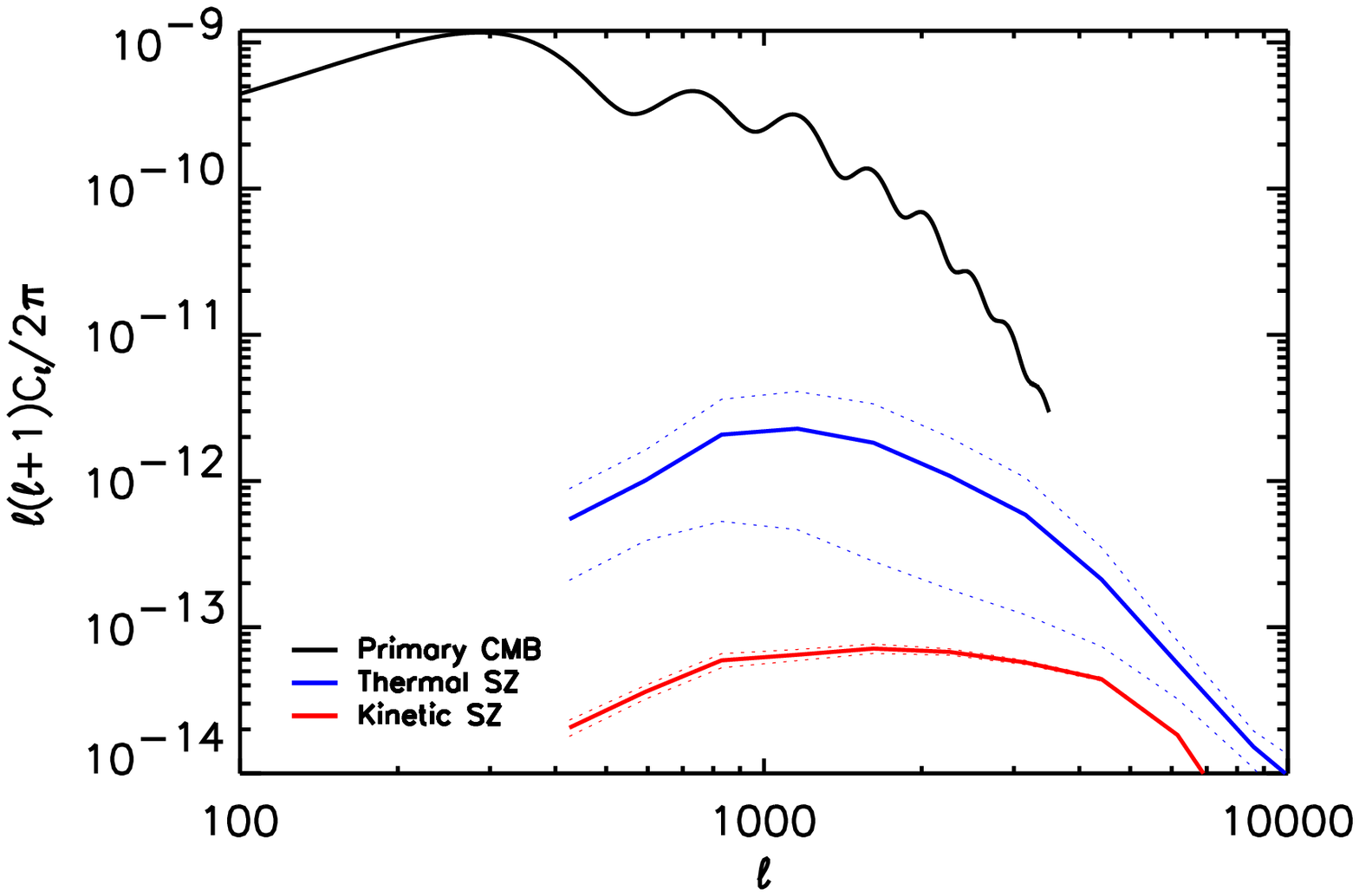}\\
\centering \leavevmode\epsfysize=6cm \epsfbox{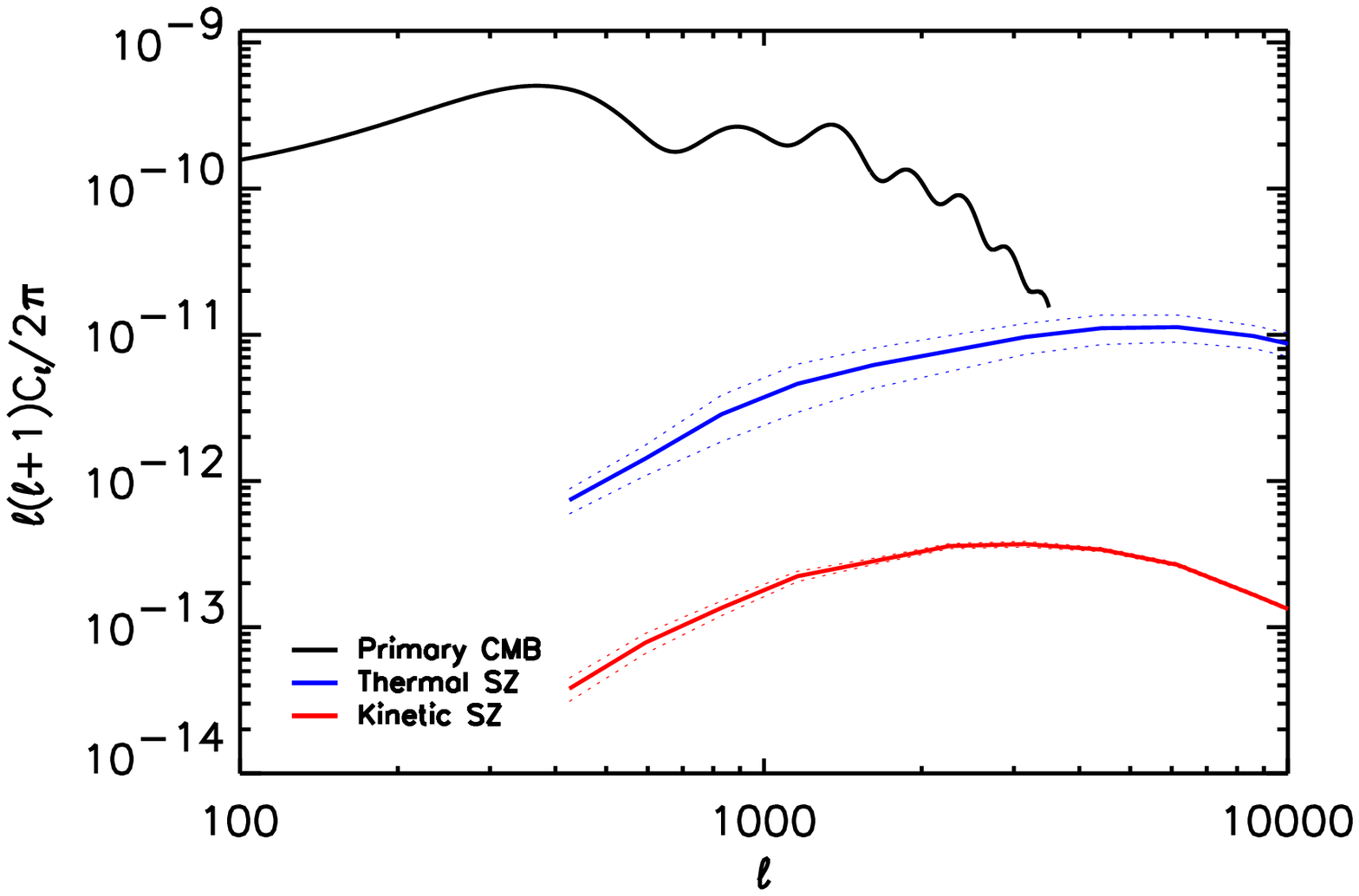}\\
\caption[cls]{\label{f:cls} In each panel, we see
the primary CMB, thermal SZ and kinetic SZ angular power spectra obtained
for each cosmology. From top to bottom we have $\Lambda $CDM, $\tau
$CDM and OCDM, respectively. The thermal spectrum is plotted for the
Rayleigh--Jeans part of the frequency spectrum, and can be scaled to
other frequencies. The dotted lines indicate $1$-sigma errors,
obtained by bootstrap resampling the thirty map realizations.}
\end{figure}

From Figure~\ref{f:cls} we see that the low-density models present 
similar shapes and amplitudes for the thermal and kinetic angular
power spectra. In the critical-density case both effects
produce
less power at all scales, which can be explained by
there being fewer objects along the line of sight
for this model. On large angular scales, the thermal $C_\ell$  
curves typically exceed the kinetic ones by a factor of about 25
for all cosmologies. On smaller scales ($\ell \ga 2000$)
the gap between the thermal and kinetic power spectrum increases
slowly in the low-density models, whereas in the $\tau $CDM model it decreases 
rapidly as 
the thermal $C_\ell$ curve falls with $\ell$. This can be understood if we
take into account that although the simulations are normalized to give
the same abundance of objects at redshift zero, in the low-density
models there are SZ sources visible to much higher redshifts. 
As we will see below, on small angular scales the thermal SZ 
spectrum receives strong contributions from sources above redshift
one. The lack of distant sources in the $\tau $CDM model is then
reflected in the thermal SZ power spectrum by a strong reduction of
power on these scales. 

In the $\Lambda $CDM cosmology the amplitude of the thermal SZ power
spectrum becomes comparable to the power spectrum of the primary CMB
around $\ell=2000$. This prediction agrees with earlier results from
numerical simulations (Refregier et al.~2000a; Seljak et al.~2000;
Springel et al.~2000) and confirms that the SZ power spectrum
should not be an important source of contamination for MAP (its
amplitude is well below the projected noise for this mission, see
Refregier et al.~2000a,b), whereas it should be detectable by
{\it Planck} and other experiments probing smaller angular scales 

\begin{figure}
\centering \leavevmode\epsfysize=6cm \epsfbox{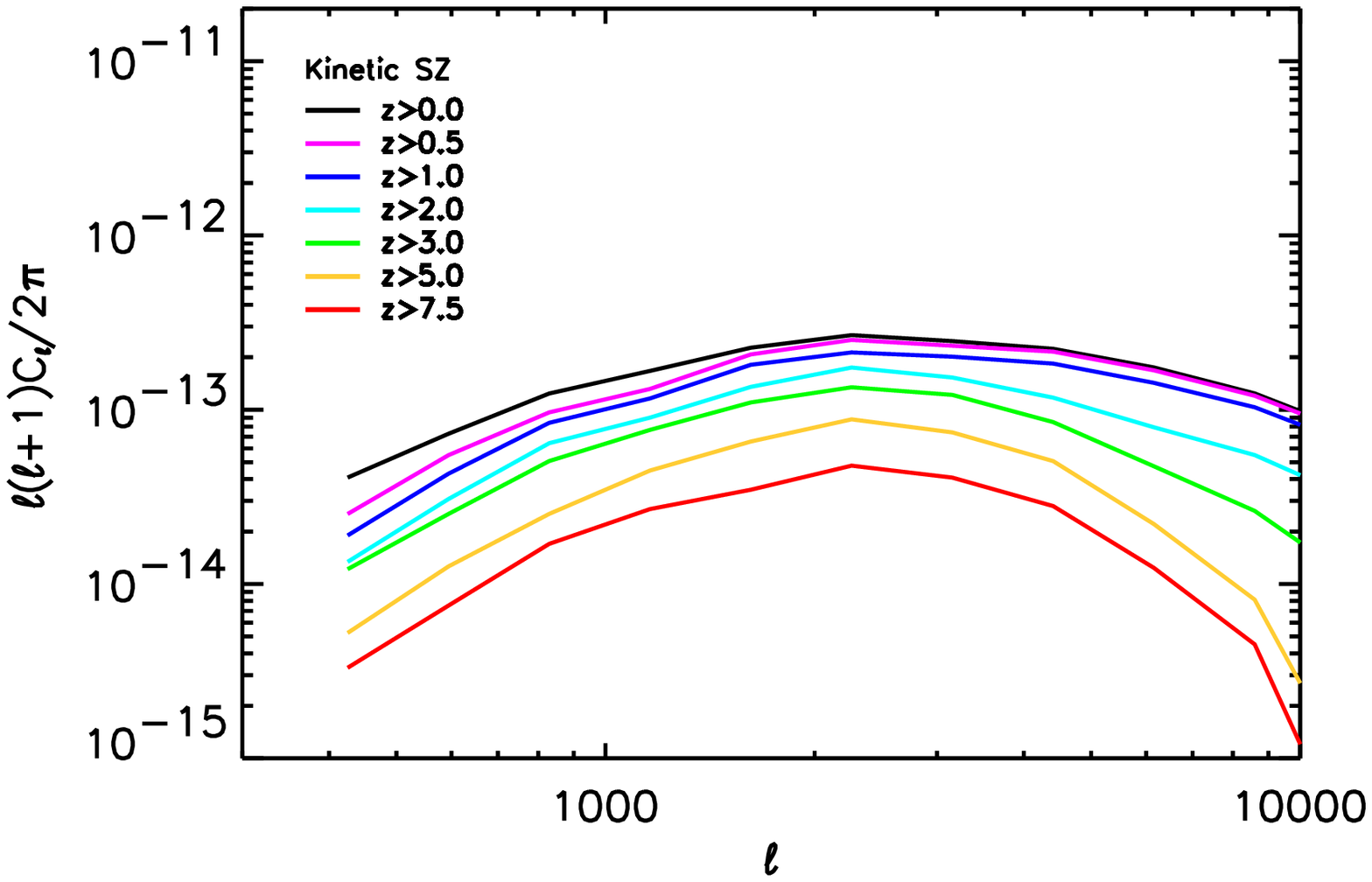}\\ 
\centering \leavevmode\epsfysize=6cm \epsfbox{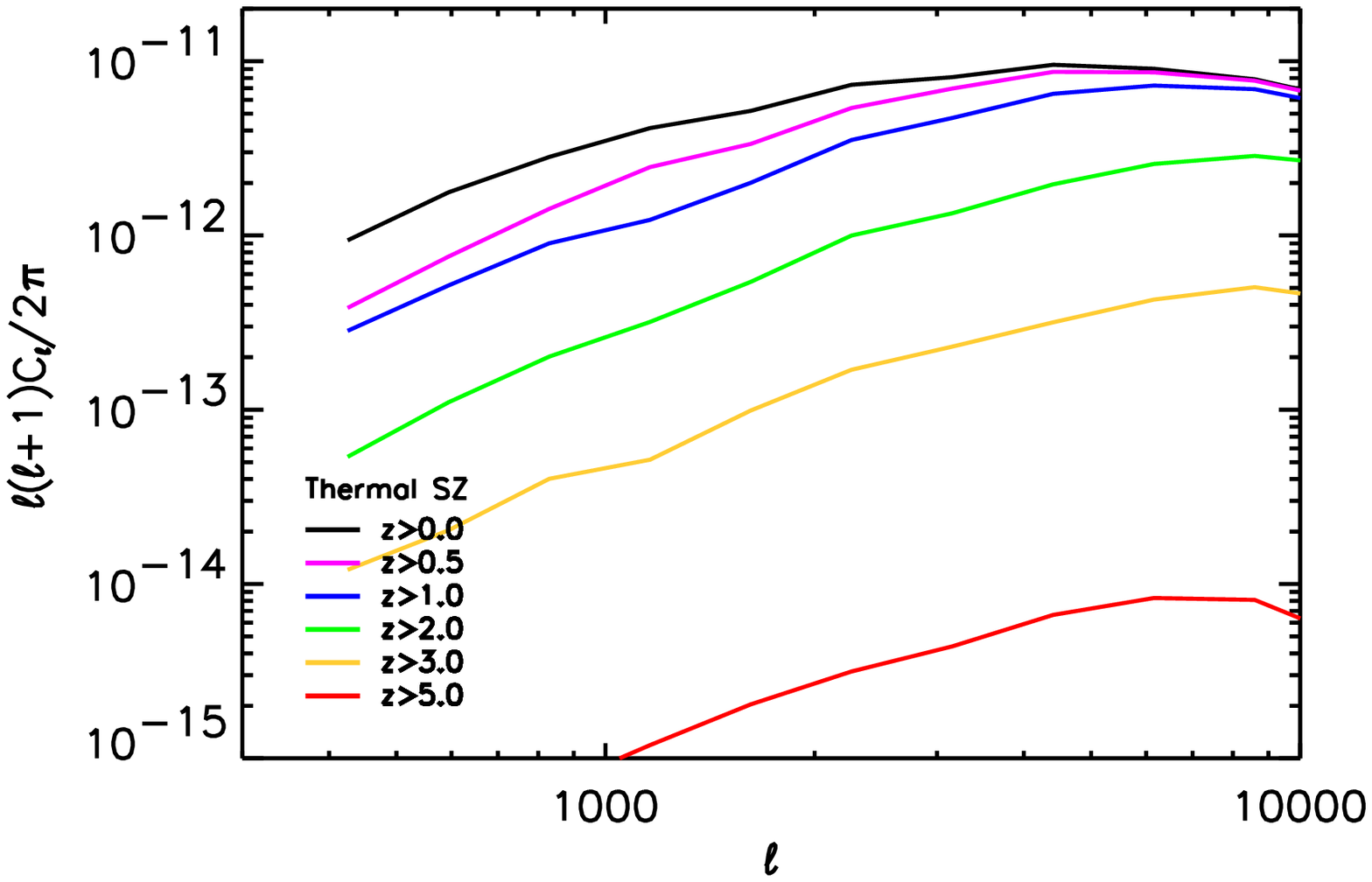}
\caption[cls_z]{\label{f:cls_z} The redshift dependence of the power
spectrum of the kinetic effect is shown in the upper panel, with the
lower panel showing the thermal effect for comparison. In each case,
the spectrum is shown for the accumulated signal from redshifts
greater than a given value. These plots refer to the $\Lambda $CDM
simulation.}
\end{figure}

Figure~\ref{f:cls_z} shows the dependence of the thermal and kinetic angular
power spectra with redshift.  We see that the kinetic effect has had a much
weaker dependence on redshift (for $\ell \simeq 2000$ about 15 per
cent of the power has origin above $z>7.5$), and indeed, while sub-dominant now,
was dominant at redshifts earlier than about three.  On large angular scales,
the thermal power spectrum is clearly dominated by sources at low redshifts.
More than two-thirds of the power on these scales is generated at redshifts
$z<1$, whereas on small scales the majority of the signal is produced above
redshift 1.  On intermediate scales there is still a reasonable fraction of
power with origin above $z=1$; for instance at $\ell=2000$ that fraction is 
about
50 per cent.

In general our results show broad agreement with the SZ power spectra obtained
from numerical simulations by Refregier et al.~(2000a), Seljak et al.~(2000) and
Springel et al.~(2000).  In the first two cases the authors have also studied
three CDM models, but made no predictions for the kinetic SZ power spectrum.  In
Springel et al.~(2000) the angular power spectrum is computed for both effects,
but results are only reported for the $\Lambda $CDM cosmology.  After taking
into account the difference between simulations on $\sigma _8$ and $\Omega _Bh$
we find that the thermal spectra agree quite well on large angular scales.
In particular, our thermal $\Lambda $CDM spectrum agrees remarkably well with
Springel et al.~(2000) for a wide range of $\ell$ (the difference is within
20 per cent for $\ell\la $1200).  This is an encouraging result as all
simulations should agree on these scales (as shown in Refregier et al.~2000a, 
the finite size of the simulation boxes has very little influence on the SZ 
power spectrum).

On small angular scales different simulations give rather different results
for the thermal effect.  Generally speaking our simulations predict less power
than Springel et al.~(2000) and more power than Refregier et al.~(2000a) and
Seljak et al.~(2000).  However, in the critical-density case our thermal curve
peaks before and falls faster than the lower limit for the spectrum reported by
Refregier et al.~(2000a).  The fact that Springel et al.~(2000) predict more 
power
on these scales is most likely reflecting the higher resolution of their
simulations.

The only available results for comparison of the kinetic spectra are those of 
Springel et al.~(2000) for the $\Lambda $CDM
cosmology. The agreement between simulations is not as
good as in the thermal case, as the shapes of the kinetic curves differ on large 
scales. Our simulations predict about $\sim$ 3.5 times less power at $\ell\simeq 
400$. However simulations show good
agreement in the range $1000\la \ell\la 2000$. 
Different assumptions concerning the ionization history of the gas may
be the main source of differences in the  spectra at very high $\ell $. 

\subsection{Kinetic effect versus thermal effect}
 
\vspace{0.5cm}
\begin{figure}
\centering \leavevmode\epsfysize=6cm \epsfxsize=8cm \epsfbox{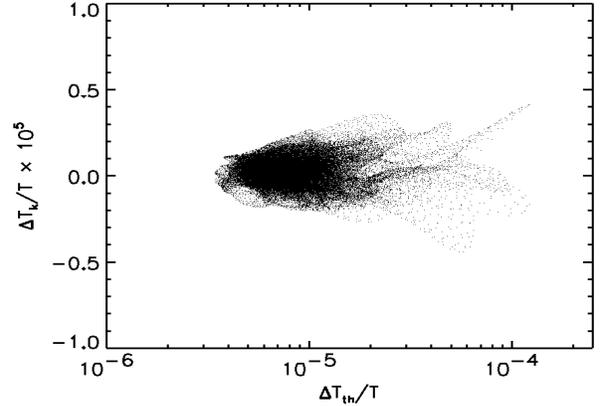}\\
\caption[scatter]{\label{f:scatter} A scatter plot of the kinetic
effect against the thermal effect, pixel-by-pixel, for a single
$\Lambda$CDM map.}
\end{figure}

Given the smallness of the kinetic effect, it is advantageous to have
an idea where on the sky to look in order to see it. Such a guide is
given by the thermal effect, which indicates locations where
gravitational collapse has concentrated scattering
electrons. Figure~\ref{f:scatter} is a pixel-by-pixel scatter plot of
the kinetic effect versus the thermal effect, found in a single map of
the $\Lambda $CDM simulation. The kinetic signal is roughly symmetric
around zero and its dispersion increases for the pixels with higher
thermal distortions.  The `tendrils' are the pixels associated with
particular large bright features in the maps.

\vspace{0.5cm}
\begin{figure}
\centering \leavevmode\epsfysize=6cm \epsfbox{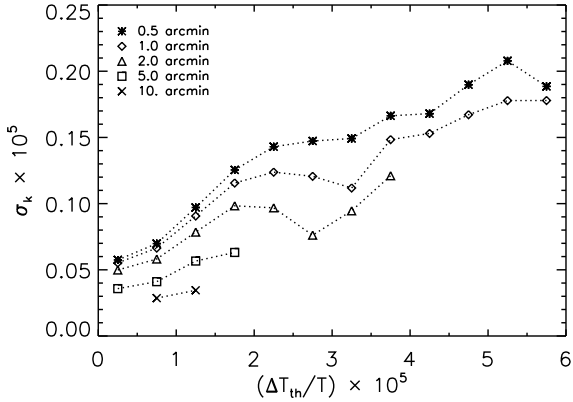}\\ 
\caption[yksigma]{\label{f:yksigma} For a given value of thermal
distortion, this plot shows the dispersion of the values for the
kinetic distortion found at the same spatial locations. For the
plotted values, the distribution is quite well approximated by a
gaussian, whereas for larger values it is dominated by nongaussian
effects.}
\end{figure}
 
An interesting question is what values of kinetic distortion are
expected at locations with a given thermal distortion, and this is
shown in Figure~\ref{f:yksigma}. We binned the pixels of all thirty
$\Lambda$CDM maps according to their thermal values and studied the
distribution of kinetic distortions. As long as the distortion is not
too large, these distributions are well approximated by gaussians of
width $\sigma _{\rm k}$ as shown in the figure, the brighter thermal
sources corresponding typically to larger kinetic distortions. This
enhancement at locations of larger thermal signal is what allows a
statistical detection of the magnitude of the kinetic effect (Haehnelt
\& Tegmark 1996; Aghanim et al.~1997) and may allow detections in some
rare bright objects. For a discussion of identifying the kinetic
effect through the component separation process, see Hobson et
al.~(1998). 

A similar exercise can be carried out for identified sources rather
than pixels. To mimic typical observational procedures, we first
subtract the mean signal from the thermal maps. We then identify
source locations in the thermal maps using the SExtractor package
(Bertin \& Arnouts 1996), and determine the total thermal flux of the
sources (at a given frequency channel) by adding up the flux in all
contiguous pixels until the pixel intensity falls below half the
maximum. In order to reduce source confusion effects we only consider
bright resolved sources which satisfy the selection criterion of
having a half maximum intensity above twice the rms
distortion in the maps. We then compute the
corresponding flux of the kinetic effect by summing the same set of
pixels in the kinetic maps. Figure~\ref{f:sscatall} shows the
resulting scatter plot, at frequency 143 GHz, made from all thirty of
the one arcminute resolution $\Lambda $CDM maps (unlike the pixel
scatter plots which are from a single map). 

\vspace{0.5cm}
\begin{figure}
\centering \leavevmode\epsfysize=6cm \epsfbox{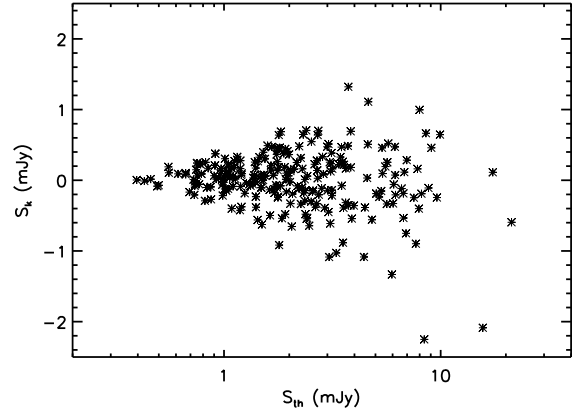}\\ 
\caption[sscatall]{\label{f:sscatall} As Figure~\ref{f:scatter}, but
now showing the total flux of sources as described in text.}
\end{figure}

As expected the largest kinetic distortions
correspond to sources with stronger thermal fluxes. However, at this frequency 
the typical flux of the kinetic effect is around one order of magnitude below 
the thermal, though for specific clusters the ratio can be as high as one third.

\section{Conclusions}

We have used hydrodynamical simulations to study a number of aspects
of the kinetic SZ effect, including its dependence on cosmological
parameters. We have studied the redshift dependence, pixel histograms,
angular power spectra, and the correlation between the kinetic and thermal
effects. We have confirmed that the kinetic effect has a dispersion
more than a factor five below 
the thermal (in the Rayleigh--Jeans region), leading to an angular power
spectrum a factor of typically twenty-five lower in the low-density
cosmologies. For critical density we
have found a smaller difference, but with much greater statistical
uncertainty. For the thermal effect the angular power spectrum is mostly 
generated at redshifts below one, while for the kinetic effect a significant 
amount of the power has origin above redshift two.
The correlation of the kinetic effect with the thermal confirms and 
quantifies the expected enhancement of the kinetic effect in regions with a 
strong thermal signal.

Our simulations are of ideal size to study the SZ effect on scales
between one and several arcminutes, which is currently a resolution
attracting great interest. Our results complement perfectly the
small-scale simulation work of Bruscoli et al.~(1999) and Gnedin \&
Jaffe (2000), and recent semi-analytic work including that of Benson
et al.~(2000) and Valageas et al.~(2000). The theory of these
secondary anisotropies is now becoming highly developed; the key
challenges for the kinetic effect lie very much on the observational
side.

\section*{Acknowledgments}

We are indebted to Hugh Couchman and Frazer Pearce for their part in
writing the {\tt Hydra}} hydrodynamical $N$-body code used to generate
the simulation data used in this work. We thank Jim Bartlett, Julian
Borrill, Ian Grivell, Andrew Jaffe, Scott Kay, Alexandre Refregier,
Volker Springel, Aprajita Verma and Martin White for helpful
discussions and comments. ACdS and DB were supported by FCT (Portugal)
and PAT by a PPARC Lecturer Fellowship.  We acknowledge use of the
Starlink computer systems at Imperial College and at Sussex. The
simulations were carried out on the BFG--HPC facility at Sussex funded
by HEFCE and SGI, and part of the data analysis on the {\sc cosmos}
National Cosmology Supercomputer funded by PPARC, HEFCE and SGI.

\bsp
\end{document}